\documentclass[%        Class options:
aps,%                   American Physical Society
prd,%                   Physical Review D
showpacs,%              Displays PACS after abstract
preprint,%              Preprint layout
tightenlines,%          Single spaced lines
nofootinbib,%           Does not treat footnotes as references
floatfix]%              Fixes float errors
{revtex4}%              REVTEX 4 Package used
\usepackage{graphicx,%  Default Latex 2eps package for embedding figures
}

\begin{document}
\title{KamLAND neutrino spectra in energy and time:\\
Indications for reactor power variations\\
and constraints on the georeactor}
%-------------------------------------------------------------
\author{G.L.\ Fogli, E.\ Lisi, A.\ Palazzo, and A.M.\ Rotunno}
\affiliation{
Dipartimento di Fisica and Sezione INFN
di Bari, Via Amendola 173, 70126, Bari, Italy}
%---------------------------------------------------------------
%\date{{\today}}
%------------------------------------------------------
\begin{abstract}
The Kamioka Liquid scintillator Anti-Neutrino Detector (KamLAND)
is sensitive to the neutrino event spectrum from (mainly Japanese) nuclear
reactors in both the energy domain and the
time domain. While the energy spectrum
of KamLAND events allows the determination of the neutrino oscillation
parameters, the time spectrum can be used to monitor  known and unknown  neutrino sources. By using available monthly-binned data on
event-by-event energies in
KamLAND and on reactor powers in Japan, we perform a likelihood  
analysis of the neutrino event spectra in energy and time, and find 
significant indications in favor of time variations of the 
known reactor sources, as compared with the hypothetical case of
constant reactor neutrino flux. 
We also find that the KamLAND data place interesting upper limits on
the power of a speculative nuclear reactor operating in the
Earth's core (the so-called georeactor); such limits are  
strengthened by including solar neutrino constraints on the neutrino
mass and mixing parameters. 
Our results corroborate the standard 
interpretation of the KamLAND signal as due to oscillating
neutrinos from known reactor sources.
\end{abstract}
 \maketitle

%%%%%%%%%%%%%%%%%%%%%%%%%%%%%%%%%%%%%%%%%%%%%%%%%%%%%%%%%%%%%%%%%%%%%%
%%%% Section I %%%%%%%%%%%%%%%%%%%%%%%%%%%%%%%%%%%%%%%%%%%%%%%%%%%%%%%
%%%%%%%%%%%%%%%%%%%%%%%%%%%%%%%%%%%%%%%%%%%%%%%%%%%%%%%%%%%%%%%%%%%%%%

\section{Introduction}

The Kamioka Liquid scintillator Anti-Neutrino Detector (KamLAND)
\cite{KamL,Grat} is sensitive to oscillations 
\cite{Pont,Maki} of reactor neutrinos \cite{Bemp}
over long baselines ($\langle L\rangle \sim 180$ km). The 
neutrino disappearance effect observed in KamLAND \cite{Kam1,Kam2,Kam3} 
provides 
an independent confirmation of the matter-enhanced adiabatic solution
\cite{Adia,Matt} 
to the solar neutrino problem \cite{Bahc,Home,SAGE,GALL,GNOx,SKso,SNO1,SNO2} 
at large mixing angle (LMA), 
with best-fit oscillation parameters 
$(\delta m^2,\sin^2\theta_{12})\simeq(7.9\times 10^{-5}\mathrm{\ eV}^2,0.31)$
\cite{Kam2,SNO2} in standard notation \cite{PDG4}. In addition, the current
KamLAND statistics and  energy resolution allow to track the oscillatory 
pattern of reactor neutrinos in the energy domain
for about half a period \cite{Kam2}.

Being a real-time detector, KamLAND can also track neutrino
source variations  in the time domain. In particular,
significant power variations of some Japanese reactors occurred during
data taking \cite{Grat}, 
leading to expected variations in the KamLAND neutrino
event rate \cite{Kam2}. The KamLAND sensitivity to time variations was estimated
to reach potentially the $\sim 2.3\sigma$ level 
through the unbinned test proposed in \cite{Unbi},
where only time information (and no energy information) was considered.

There is also, in principle, an interesting interplay between time and 
energy information
in KamLAND. In the presence of neutrino
oscillations, time variations of reactors 
placed at different distances produce time variations of the energy spectrum.
Figure~1 shows, e.g., that by ``switching off'' one of the most powerful
nuclear reactor plants in Japan (namely, Kashiwazaki) 
one gets not only an overall decrease
of the spectrum normalization, but also a slight displacement of
the dip in the oscillatory pattern. In the same figure, one can also 
see the effect of a hypothetical reactor at the center of the Earth
(the so-called georeactor \cite{Geo1}), which would increase the KamLAND
spectrum by a factor which is constant in time but, in general,
 not uniform in energy.
A joint analysis in the energy and time domain would be appropriate
to study such effects.

So far, the KamLAND collaboration has published only one   
test of the time-variation hypothesis, which makes use of a relatively
coarse time binning and of no energy information.
The results are shown in the first figure of Ref.~\cite{Kam2}, where the observed event rates---grouped 
in five data points---are plotted against the unoscillated reactor 
neutrino flux, and a positive (expected) correlation is seen
to emerge. However, the statistical
difference between the two extreme cases in this test
(with and without time variations of the neutrino flux) 
is only $\Delta\chi^2=3.3$ \cite{Grat}, i.e., smaller
than $2\sigma$. At a similar significance level, 
the extrapolation to ``zero reactor power'' is
consistent with the known 
background, but yields poor constraints on possible unidentified sources such
as the georeactor \cite{Kam2}.

The power of a time-variation test---as the one described in 
\cite{Kam2}---can 
be improved by exploiting additional information. For instance, 
daily  data about individual Japanese reactor operations
are available to the KamLAND collaboration, through an agreement 
with the power companies \cite{Ku03}.  In principle, these data allow one to perform
detailed likelihood analyses of the event spectra 
not only in the energy domain 
(as those, e.g., in \cite{Kam1,Kam2,SKso,SNO2,Fo03,Ia03,Schw,Ricc,Smir,Ba04,Va04,Al04,Ma05,Go05,St05}) 
but also in the time domain, thus providing
statistically more powerful tests of reactor power variations and 
of the georeactor hypothesis. Unfortunately, 
daily reactor data are classified \cite{Ku03}.

Recently, monthly-binned data from 
nuclear reactors and from KamLAND have become publicly 
available. In particular, average Japanese reactor powers in each calendar
month can be found at \cite{JAIF}. The sequence of published KamLAND events
\cite{Kam3} in monthly bins, 
together with the corresponding detector livetime (in seconds), can be found 
in \cite{File}. The availability of these
data has prompted us to extend the likelihood analysis of 
KamLAND data in the energy domain (event by event) \cite{Kam1,Kam2,Schw} 
so as to include the 
time domain (monthly binned). We find that the joint
maximum-likelihood analysis in energy and time can provide a significant 
($\sim 3\sigma$) indication
in favor of time variations of reactor powers,
as compared with the case of average constant powers.
In addition, we find no 
indication in favor of a georeactor contribution, and we set upper bounds
on its power. In both cases, we discuss the role of additional
solar neutrino constraints on $(\delta m^2,\sin^2\theta_{12})$.
Our results corroborate the standard interpretation
of the KamLAND signal as due to flavor oscillations of
neutrinos coming from known reactor sources.

The structure of this paper is as follows. In Sec.~II we reproduce, as 
a preliminary but relevant check, the official KamLAND unbinned
likelihood analysis in the energy domain 
\cite{Kam2,Kam3}. In sec.~III we extend the analysis to the (monthly
binned) time domain, and show that significant indications in favor of 
reactor time variations emerge from the data.
In Sec.~IV we discuss the effects of a  hypothetical georeactor, and set
upper bounds on its power. We summarize our results in Sec.~V.

A final remark is in order. Our results, although encouraging, cannot---and
must not---be taken as a substitute for future, official KamLAND
tests of hypotheses about the reactor sources. In fact, as described
in the following, our approach involves some unavoidable approximations, which
could be easily removed by the KamLAND
collaboration---possibly leading to somewhat different results.
Nevertheless, we think
that our approximate analysis in the energy-time domain
may represent an interesting
step beyond previous KamLAND data analyses, where the  time
information is absent.

\section{Likelihood analysis in energy}

The KamLAND experiment has collected so far $N_\mathrm{obs}=258$ events
in a fiducial mass $M=543.7$ tons, during a total livetime 
$\Delta t=515.1$ days \cite{Kam2}:
%.................................................................
\begin{equation}
M\cdot\Delta t=0.766 \mathrm{\ kTy}.
\end{equation}
%................................................................
Details on the likelihood analysis 
of the energy spectrum of such events are available at \cite{Kam3}. 
In this Section we reproduce
the results of the official KamLAND likelihood analysis in energy
\cite{Kam2}, before generalizing it to the time
domain in Sec.~III.

In general, the KamLAND unbinned likelihood 
function $\mathcal{L}$ can be written as \cite{Kam1,Kam2,Schw}:
%...................................................................
\begin{equation}
\label{L}
\mathcal{L}=\mathcal{L}_\mathrm{rate}\times
\mathcal{L}_\mathrm{shape}\times \mathcal{L}_\mathrm{syst}\ , 
\end{equation}
%.....................................................................
where the three factors embed information on the total event rate,
on the spectrum shape, and on systematic uncertainties. The evaluation
of $\mathcal{L}$ implies a detailed calculation of the absolute
spectrum of events (signal plus background), whose ingredients are
briefly described below.

\subsection{Reactor input}

The reactor signal in KamLAND
is essentially
generated by 20 nuclear reactor power plants (16 in Japan and
4 in Korea) located at different distances $L_j$ and 
characterized by 
different thermal powers $P_j^\mathrm{th}$ \cite{Grat}. For Japanese reactors ($j=1,\dots,16$),
the sequence of monthly-averaged thermal 
powers $P_{jm}^\mathrm{th}$ 
(where $m$ is a monthly index) can be recovered
from the corresponding sequence of average electric powers $P_{jm}^\mathrm{el}$
available in \cite{JAIF}, by using the relation  
$P^\mathrm{th}\simeq 3\, P^\mathrm{el}$ \cite{Prop}. The time interval of interest
for the current KamLAND analysis spans $m=1,\dots,23$ months, from
March 2002 to January 2004 included \cite{Kam2,File}. 
In each month, the KamLAND
detector livetime $\Delta t_m$ (with $\sum_m\Delta t_m=\Delta t$) is given
in \cite{File}. The average thermal
power of the $j$-th Japanese reactor
during the total KamLAND livetime $\Delta t$ can thus be 
approximately estimated as
%...................................................................
\begin{equation}
P^\mathrm{th}_j\simeq\frac{\sum_{m=1}^{23}
P^\mathrm{th}_{jm} \Delta t_m}{\Delta t}\ ,
\end{equation}
%.....................................................................
where we are implicitly neglecting variations of the
reactor powers (and of the detector livetime) over time scales 
shorter than a month.%
%----------
\footnote{
This approximation could be
removed by using, e.g., daily data, which are available only within
the KamLAND collaboration.}
%----------------
We have not found monthly information about the four 
Korean reactor plants ($j=17,\dots, 20$), which we simply  
assume to have constant powers 
($P^\mathrm{th}_{jm}=P^\mathrm{th}_j$), where $P^\mathrm{th}_j$
is taken as a typical fraction (80\%) of the nominal 
thermal power quoted in 
\cite{Grat}.%
%-----------------------------
\footnote{This is a minor approximation, since Korean reactors contribute only $\sim 3\%$ to the KamLAND signal.}
%---------------------------
For all reactors, the average fuel components $q_f$ ($f=1,\dots, 4$) 
are taken as \cite{Kam2}
%...................................................................
\begin{equation}
^{235}\mathrm{U}:\,
^{238}\mathrm{U}:\,
^{239}\mathrm{Pu}:\,
^{241}\mathrm{Pu}=0.563:0.079:0.301:0.057
\end{equation}
%.....................................................................
at all times, with average fission energies $E_f= 201.7$, 210.0, 205.0, 
and 212.4 MeV, respectively \cite{Boeh}. 
We do not have enough information to implement 
fuel burn-up corrections \cite{Kam2,Mura} to individual reactors. 

Within the above approximations, 
the time-averaged 
differential neutrino flux at KamLAND (number of neutrinos per unit 
of time, area, and energy) 
is then given by \cite{Bemp}
%...................................................................
\begin{equation}
\label{nuflux}
\frac{d\phi}{dE_\nu}\simeq \sum_{j=1}^{20}\sum_{f=1}^4\frac{P^\mathrm{th}_j}{4\pi L_j^2}
\frac{q_f}{E_f}\frac{dN_f}{d E_\nu}\ ,
\end{equation}
%.....................................................................
where we assume, for the $f$-th spectral component, the 
parametrization \cite{Voge}
%...................................................................
\begin{equation}
\frac{dN_f}{dE_\nu}=\exp(a_0^f+a_1^fE_\nu+a_2^kE_\nu^2)\ ,
\end{equation}
%.....................................................................
the $a_h^f$ coefficients being reported in \cite{Voge}. In the presence of oscillations,
each $j$-th reactor
term in Eq.~(\ref{nuflux}) must be multiplied by the 
corresponding neutrino survival probability
$P_{ee}(E_\nu,L_j)$.

We have made two reassuring checks of the above reactor power input.
As a first check, we have estimated the total integrated thermal power flux
over the detector livetime,
%...................................................................
\begin{equation}
\sum_{j,m}\frac{P_{jm}^\mathrm{th}\Delta t_m}{4\pi L^2_j}=
697\mathrm{\ J/cm}^2\ , 
\end{equation}
%.....................................................................
in good agreeement  
with the official KamLAND value of 701 J/cm$^2$ \cite{Kam2}. 

As a second check, we have calculated the so-called integrated
fission number flux \cite{Kam3},
%...................................................................
\begin{equation}
	\sum_{j,f}\frac{q_f}{E_f}\frac{P_j^\mathrm{th}\Delta t}{4\pi L^2_j}\ ,
\end{equation}
%.....................................................................
and its (binned) distribution over the reactor distance
$L$. Figure~2 shows that our results are very close
to the official KamLAND ones, as taken from \cite{Kam3}.

\subsection{Detection input}

Given the differential neutrino flux in Eq.~(\ref{nuflux}), the 
time-averaged energy spectrum of reactor events in KamLAND (number of 
expected events per
unit of prompt positron energy $E$) is given by
%...................................................................
\begin{equation}
\label{SE}
S(E)=\varepsilon \, n\, M\,\Delta t
\int dE_\nu\, \frac{d\phi}{d E_\nu}\int dE'\, \frac{d\sigma(E_\nu,E')}{dE'}\,r(E,E')\ ,
\end{equation}
%.....................................................................
where $\varepsilon=0.898$ is the overall efficiency (after all cuts 
\cite{Kam2}), 
$n$ is the target density ($0.848\times 10^{29}$ protons/ton) \cite{Kam2},
$r(E,E')$ is the energy resolution function (with 
Gaussian width equal to $7\%(E'/\mathrm{MeV})^{1/2}$) \cite{Kam3}, 
and $\sigma$ is the
inverse beta decay cross section, estimated as
%...................................................................
\begin{equation}
\frac{d\sigma(E_\nu,E')}{dE'}\simeq \sigma(E_\nu)\,
\delta(E_\nu-E'-0.782~\mathrm{MeV})\ ,
\end{equation}
%.....................................................................
with $\sigma(E_\nu)$ taken from \cite{Beac}. In $r(E,E')$,
we allow for a systematic
offset of the prompt (true) energy scale,
%.....................
\begin{equation}
\label{scale}
E'\to E'(1+\alpha) \ ,
\end{equation}
%.....................
with standard deviation
$\sigma_\alpha=2\times 10^{-2}$ \cite{Kam2}.

Above the current analysis threshold 
($E_\mathrm{thr}=2.6$~MeV), we estimate a total
of 377.3 reactor events in the absence of oscillations. This value 
is about $3\%$ higher
than the official KamLAND estimate (365.2 events 
\cite{Kam2}); we obtain a $+3\%$ difference also in comparison with older data
\cite{Kam1} (89.7 events against the official 86.8 estimate \cite{Kam1}). We have not been
able to trace the source of this modest systematic difference, which we choose 
to compensate ``ad hoc'' in the following, 
through a fudge factor $f=0.97$ multiplying the right hand side 
of Eq.~(\ref{SE}).%
%-------------------------------
\footnote{This small adjustment ($3\%$) is only $\sim 1/2$ of the
KamLAND normalization error ($6.5\%$). Removal of such adjustment 
does not appreciably change any of our results.}
%--------------------------------

Finally, one must consider the background energy spectrum $B(E)$
expected over the livetime $\Delta t$. 
This spectrum has three main components, as described  in detail in 
\cite{Kam2,Kam3}: the accidental background
$B_1$, the $^8$He-$^9$Li background $B_2$, and the $^{13}$C$(\alpha,n)^{16}$O
background $B_3$. While $B_1$ and $B_2$ can be estimated with very small
uncertainties \cite{Kam2} (that we set to zero), the normalization of
the third background is
poorly known in both its low-energy ($<5.4$ MeV) and high-energy 
($>5.4$ MeV) components \cite{Kam3} 
($B_3'$ and $B_3''$, respectively). We then assume free normalization 
factors ($\alpha'$ and $\alpha''$) for such components. In
conclusion, we take the absolute background spectrum as
%...................................................................
\begin{equation}
B(E)=B_1(E)+B_2(E)+\alpha'B_3'(E)+\alpha''B_3''(E)\ , 
\end{equation}
%.....................................................................
where the $B_1$, $B_2$, $B_3'$ and $B_3''$ components 
are taken from \cite{Kam3}, while 
$\alpha'$ and $\alpha''$ are free (positive) parameters.

\subsection{Likelihood function and oscillation parameters}

The absolute energy spectrum of events expected above the analysis
threshold can always be factorized into the total number
of events $N_\mathrm{theo}$ times the probability distribution 
in energy $D(E)$, namely
%...................................................................
\begin{equation}
S(E)+B(E)=N_\mathrm{theo}\cdot D(E)
\end{equation}
%.....................................................................
with
%...................................................................
\begin{equation}
\int_{E_\mathrm{thr}}dE\, D(E)=1\ .
\end{equation}
%.....................................................................
We remind that both $N_\mathrm{theo}$
and $D(E)$ depend on the systematic energy offset $\alpha$, as
well as on the free background parameters $\alpha'$ and $\alpha''$.
In the presence of oscillations, they also depend on the 
the mass-mixing parameters 
$(\delta m^2,\sin^2\theta_{12})$.%
%-----------------------------------
\footnote{In this work we do not consider subleading
three-neutrino oscillation effects, i.e., we assume $\theta_{13}=0$
in standard notation. Within current bounds ($\sin^2\theta_{13}
\lesssim\mathrm{few}\%$) we do not expect this approximation to be crucial.
We also neglect small Earth matter effects
on reactor neutrino propagation \cite{Ba04}. }
%------------------------------------

Given the previous definitions, the
first likelihood factor in Eq.~(\ref{L}) can be written as
(see also \cite{Schw}):
%...................................................................
\begin{equation}
\label{Lr}
\mathcal{L}_\mathrm{rate}=\frac{1}{\sqrt{2\pi}\sigma_\mathrm{rate}}
\exp\left[
-\frac{1}{2}\left(
\frac{N_\mathrm{theo}(\delta m^2,\,\sin^2\theta_{12}\,;\,
\alpha,\,\alpha',\,\alpha'')-N_\mathrm{obs}}{\sigma_\mathrm{rate}}
\right)^2
\right]
\end{equation}
%.....................................................................
where $N_\mathrm{obs}=258$ is the total number of observed events 
\cite{Kam2},
and the total error is the sum of the statistical and systematic
($s=6.5\%$ \cite{Kam2,Kam3}) uncertainties, 
%...................................................................
\begin{equation}
\sigma_\mathrm{rate}^2=N_\mathrm{theo}+
(s\,N_\mathrm{theo})^2\ .
\end{equation}
%.....................................................................
The second likelihood factor
in Eq.~(\ref{L}) is the product of the probability
that the $i$-th event ($i=1,\dots,N_\mathrm{obs}$) occurs with the
observed energy $E_i$,
%...................................................................
\begin{equation}
\mathcal{L}_\mathrm{shape}=\prod_{i=1}^\mathrm{258}
D(E_i\,|\,\delta m^2,\,\sin^2\theta_{12}\,;\,\alpha,\,\alpha',\,\alpha'')\ ,
\end{equation}
%.....................................................................
where the energy set $\{E_i\}$ is given in \cite{Kam3}.  
The third and last likelihood factor in Eq.~(\ref{L}) embeds the penalty
for the systematic offset $\alpha$ in Eq.~(\ref{scale}), 
%.....................................................................
\begin{equation}
\label{Ls}
\mathcal{L}_\mathrm{syst}=\frac{1}{\sqrt{2\pi}\sigma_\alpha}
\exp\left[{-\frac{1}{2}\left(\frac{\alpha}{\sigma_\alpha}\right)^2}
\right]\ .
\end{equation}
%.....................................................................
In general, further penalty factors could account for
additional KamLAND systematics (not included here for lack of 
detailed published information).

Finally, the standard $\chi^2$ function is obtained as
%.....................................................................
\begin{equation}
\chi^2(\delta m^2,\sin^2\theta_{12})=-2 \ln
\max_{\{\alpha,\alpha',\alpha''\}} 
\mathcal{L}(\delta m^2,\,\sin^2\theta_{12}\,;\,\alpha,\,\alpha',\,\alpha'')\ .
\end{equation}
%.....................................................................
Bounds on the oscillation parameters can be found by plotting isolines
of the function
%.....................................................................
\begin{equation}
	\Delta\chi^2=\chi^2-\min_{\{\delta m^2,s^2_{12}\}}\chi^2\ .
\end{equation}
%.....................................................................
The values $\Delta\chi^2=4.61$, 5.99, 9.21, and 11.83 correspond to
90, 95, 99, and 99.73\% C.L.\ for two degrees of freedom.

Figure~3 shows the bounds on the oscillation parameters from our
likelihood analysis of the KamLAND energy spectrum. The confidence
level isolines are
in very good agreement with the official ones reported in Fig.~4(a)
of \cite{Kam2}, modulo the different scales chosen for the axes.%
%-----------------------------
\footnote{We prefer to plot the---currently small---allowed regions 
in linear scale, rather than in logarithmic scale. In particular, the
log-scale in $\tan^2\theta_{12}$, introduced
in \cite{Mont} and used in \cite{Kam2}, can be usefully replaced by a linear
scale in $\sin^2\theta_{12}$, which 
preserves the $\theta_{12}$ octant symmetry \cite{Mont} when applicable 
(this is not
the case for a linear scale in $\tan^2\theta_{12}$, as used, e.g., in 
\cite{SNO2}).}
%-------------------------------
These results, together with the previous checks in this Section, 
demonstrate that we can reproduce, to a good accuracy, both the input and the output of the official 
KamLAND likelihood analysis of the energy spectrum. This check 
is also relevant to appreciate, in the next Section, the 
(small) differences induced by including the time information in the likelihood analysis.

\section{Likelihood analysis in energy and time}

The information reported in \cite{File} 
allows to separate the global KamLAND set 
of 258 event-by-event energies into 23 monthly subsets $I_m$,
%.....................................................................
\begin{equation}
\{E_i\}_{i=1,\dots,258}=\bigcup_{m=1,\dots,23}\{E_i\}_{i\in I_m}\ ,
\end{equation}
%.....................................................................
with corresponding detector livetimes $\Delta t_m$.
The goal of this section is to include such time information, together
with the set of monthly thermal reactor powers
$\{P_{jm}^\mathrm{th}\}$, into a maximum
likelihood analysis. The generalization is straightforward:
monthly neutrino fluxes $\phi_m$, signal spectra $S_m$, background 
spectra $B_m$,%
%----------
\footnote{We assume that all background components are constant
in time.}
%-------- 
and probability distributions $D_m$ are defined as
%..................................................................
\begin{equation}
\frac{d\phi_m}{dE_\nu}\simeq \sum_{j=1}^{20}\sum_{f=1}^4\frac{P^\mathrm{th}_{jm}}{4\pi L_j^2}
\frac{q_f}{E_f}\frac{dN_f}{d E_\nu}\ ,
\end{equation}
%..................................................................
%..................................................................
\begin{equation}
S_m(E)=\varepsilon \, n\, M\,\Delta t_m
\int dE_\nu\, \frac{d\phi_m}{d E_\nu}\int dE'\, \frac{d\sigma(E_\nu,E')}{dE'}\,r(E,E')\ ,
\end{equation}
%..................................................................
%..................................................................
\begin{equation}
B_m(E)=B(E)\, \frac{\Delta t_m}{\Delta t}\ ,
\end{equation}
%..................................................................
%..................................................................
\begin{equation}
S_m(E)+B_m(E)=N_\mathrm{theo}\,D_m(E)\ ,
\end{equation}
%..................................................................
respectively, fulfilling the relations
%..................................................................
\begin{eqnarray}
\sum_m S_m(E)&=&S(E)\ ,\\
\sum_m B_m(E)&=&B(E)\ ,
\end{eqnarray}
%..................................................................
and the probability normalization condition
%..................................................................
\begin{equation}
\sum_m\int_{E_\mathrm{thr}}dE\, D_m(E)=1\ .
\end{equation}
%..................................................................
The likelihood of the spectral shape information
acquires then an explicit (monthly)
time dependence, 
%..................................................................
\begin{equation}
\mathcal{L}_\mathrm{shape}=\prod_{m}\,\prod_{i\in I_m}D_m(E_i)\ ,
\end{equation}
%..................................................................
while the functional forms of $\mathcal{L}_\mathrm{rate}$ 
and $\mathcal{L}_\mathrm{syst}$ remain the same as in 
Eqs.~(\ref{Lr}) and (\ref{Ls}),
respectively. We have thus all the ingredients  to calculate a likelihood
function in energy (event-by-event) and time (monthly-binned). 

Notice that the likelihood
function in energy and time reduces to the energy-only
likelihood function in the limit of constant reactor powers $(P_{jm}^\mathrm{th}\equiv P_j^\mathrm{th})$,
up to an irrelevant overall factor (the product of 
$\Delta t_m/\Delta t$ ratios); this limit provides a useful cross-check of the numerical results.

\subsection{Constraints on the oscillation parameters}

We start the discussion of the time-dependent effects in a case
where they are (currently) not expected to play a significant
role, namely,  in the determination of the oscillation parameters
$(\delta m^2,\sin^2\theta_{12})$. The KamLAND
bounds on these two parameters are
basically dominated by two different pieces of information:
the energy spectrum shape and its normalization. In particular, the
$\delta m^2$ parameter governs the oscillation phase, which
is strongly constrained by the observation of half-period of
oscillations \cite{Kam2,St05}. 
This observation is still dominated by statistical
errors \cite{Berg}, which currently hide subleading
time-dependent effects, such as a possible shift of the ``oscillation dip''
for strong reactor power variations (as shown in
 Fig.~1). On the other hand, the $\sin^2\theta_{12}$
parameter governs the oscillation amplitude, whose bounds are
dominated by normalization systematics \cite{Berg}, which are not reduced
by adding time information. Therefore, within current uncertainties,
we do not expect  the mass-mixing bounds from the energy
spectrum analysis (Fig.~3) to be significantly changed by adding
time information.

Figure~4 shows the results of our likelihood analysis in energy and time,
which confirms the above expectations. A comparison with Fig.~3 reveals
appreciable changes only in the ``high-$\delta m^2$'' allowed region
(so-called LMA-II solution \cite{Fo03}), which  
appears to be slightly more disfavored by adding
time information. This trend allows to exclude with more confidence the
 LMA-II solution in combination with solar data (which, 
by themselves, still allow relatively high values of $\delta m^2$ 
\cite{SNO2}).
For the sake of completeness, and for later purposes, we show
in Fig.~5 the oscillation parameter bounds from 
our analysis of all current solar neutrino
data \cite{Prog} (including the latest full SNO spectral results 
\cite{SNO2}) plus the KamLAND
likelihood analysis in time and energy. The bounds in
Fig.~5 contain, to our knowledge, the largest
amount of solar and reactor neutrino
information which is publicly available at present.

\subsection{Probing time variations of the reactor neutrino flux}

Variations of the reactor powers and of the livetime efficiency
generate time variations of the event rate in KamLAND. Therefore, 
theoretical event rates including (not including) time information are expected
to track more (less) faithfully the observed event rates.
In Fig.~6  we plot the observed monthly counts in KamLAND, with respect
to our calculated counts,% 
%-----------------------
\footnote{Theoretical estimates refer to the solar+KamLAND
best-fit oscillation parameters in Fig.~5.}
%------------------------
with and without reactor power variations. The comparison of the two
panels shows at a glance that the correlation among the 23 points 
is more evident when monthly reactor powers
$(P_{jm}^\mathrm{theo})$ are included, with respect to the 
hypothetical case of constant reactor powers ($P_{jm}^\mathrm{th}\equiv P_{j}^\mathrm{th})$. Quantitatively, the correlation index decreases from 0.73 
(left panel) to 0.58 (right panel).
In the right panel, the correlation
would be further reduced for hypothetically constant detector livetimes
$(\Delta t_m=\Delta t/23)$, since all 
points would then
collapse onto a single vertical line (not shown). 

The significant
covariance between observed and calculated counts---when time information
is fully included---suggests that KamLAND is indeed tracking
reactor neutrino flux variations. In this sense, Figure~6 
qualitatively agrees
with the correlation test shown in the first figure of \cite{Kam2}.
We refrain, however, from fitting a ``straight line'' through
the points in the left panel of
Fig.~6, since we know of no clear way to include 
the point-by-point systematics and the large
statistical fluctuations in such a linear fit. A maximum-likelihood 
test of time variations appears to be
more appropriate, both to deal with small monthly counts and to
include event-by-event energies and systematics.

In order to test the null hypothesis of no time variations against the 
hypothesis of actual time variations of
the reactor neutrino flux, we introduce an auxiliary 
variable $\eta$, interpolating between the two cases. In this
way, the hypothesis test is transformed into a parameter estimation test
\cite{Lyon}.
Formally, we assume that parameter $\eta$ modulates all 
reactor powers through the equation
%..................................................
\begin{equation}
\tilde{P}_{jm}^\mathrm{th}
(\eta)=P_j^\mathrm{th}+\eta\,\Delta P_{jm}^\mathrm{th}\ ,
\end{equation}
%.......................................................
where 
$\Delta P_{jm}^\mathrm{th}
=P_{jm}^\mathrm{th}-P_j^\mathrm{th}$ are the actual power variations
in each month. Thus $\eta$ continuously ``switches on''
reactor neutrino flux variations from the 
null case $\eta=0$ (no time variations)
to the real case $\eta=1$ (actual time variations).

By using reactors powers $\tilde{P}_{jm}^\mathrm{th}
(\eta)$  
defined as in the above equation, we build
a likelihood function in energy and time 
$\mathcal{L}(\delta m^2,\sin^2\theta_{12},\eta)$, and marginalize it 
with respect to the oscillation parameters.
The results are shown in Fig.~7, in terms of the function $\Delta\chi^2(\eta)$.
The hypothetical
case of constant averaged reactor powers $(\eta=0)$
is definitely disfavored by KamLAND data, as compared with any
case including time variations  ($0<\eta<1$). In particular, the difference
with respect to 
the case of actual time variations ($\eta=1$) amounts to about $3\sigma$
($\Delta\chi^2\simeq 9$). We conclude that the results in
Fig.~7 (and, to some extent, in Fig.~6) 
can be taken as a statistically significant
indication that 
reactor neutrino flux variations have been seen in KamLAND.

Finally, it is interesting to note that,
in Fig.~7,  the addition of solar neutrino
information (through an additional $\Delta\chi^2$ function 
which depends on $(\delta m^2,\sin^2\theta_{12})$ but not on $\eta$) does not significantly change the overall bounds on $\eta$. 
In other words, as also 
observed in the previous subsection, energy and time information
are largely decoupled in KamLAND (at present).  The energy information
indicates nonzero oscillation parameters, while the time information
indicates nonzero  variations of the reactor signal rate, with no 
appreciable cross-talk between these two pieces of information.
Only with much smaller errors one might hope to see
mixed effects (e.g., time-dependent changes of the energy spectrum dip).
However, as we shall see in the next section, such ``decoupling'' of 
the  oscillation parameters 
is not necessarily preserved in nonstandard cases, e.g., in a scenario 
with a hypothetical georeactor.

\section{Constraints on the georeactor}

It has been proposed \cite{Geo1} that there could be enough
Uranium in the Earth's core to naturally start a nuclear fission chain
over geological timescales, 
with a typical power (at the current epoch)
of 3--10 TW \cite{Geo1}, and possibly
up to $\sim 30$ TW \cite{Geo2}. The latter value is probably too high to be
credible, since the addition of a typical radiogenic contribution 
of $\sim 20$ TW \cite{Fior} (not to count other sources \cite{Ande})
would exceed the total Earth heat flux (estimated to be 
$\sim 44$ TW in \cite{Poll} and recently revised down to 
$\sim 31$ TW in \cite{Hofm}). A georeactor power of $\sim 10$ TW
is, however, comparable to the global Earth heat flux uncertainty
\cite{Ande}, and thus cannot be currently
excluded by energy-budget arguments. On the other
hand, there are independent geochemical and geophysical arguments 
which seem to disfavor any significant Uranium content in the core 
\cite{McDo}.  
Despite being largely ignored
in the Earth science literature, the georeactor hypothesis
has attracted 
some attention in the particle physics literature \cite{Geos,APSR}.

In the KamLAND data analysis, a hypothetical georeactor can induce
several effects. First, it increases the overall expected 
event rate. Second, it distorts the spectrum shape, both because
its natural fuel composition can be significantly different from that of
man-made reactors, and because the oscillation phase  
for $L=R_\oplus$ is different than for $L\sim O(100)$ km.
Third, the georeactor signal is constant, while man-made reactors
induce, in general, a variable signal in KamLAND. 
Therefore, we expect that a maximum likelihood analysis of the KamLAND data
in energy and time,  including the bounds
on the oscillation parameters from solar
neutrino data, can provide interesting constraints of the georeactor
hypothesis.
Technically, we implement the georeactor hypothesis by adding
(in the KamLAND data analysis)  a 21-th
reactor at $L=6400$ km,%
%---------------------------------
\footnote{The georeactor radius is $\Delta L\lesssim O(10)$ km \cite{Geo2} 
and thus negligible in this context ($\Delta L/L\lesssim 10^{-3}
\ll \Delta E/E$).}
%-----------------------------------
 with arbitrary constant power $P_\mathrm{geo}$.
For definiteness, we assume a current georeactor fuel ratio
$^{235}\mathrm{U}:{}^{238}\mathrm{U}\simeq 0.75:0.25$,
with no significant Pu contribution \cite{Geo2}.

Figure~8 shows the bounds on the oscillation parameters
from our KamLAND maximum-likelihood analysis
in energy and time, for the illustrative case $P_{\mathrm{geo}}=15$ TW.
The ``wavy'' contours of the lowest-$\delta m^2$ allowed
region in Fig.~8 reflect the
``ripples'' created by georeactor neutrino oscillations 
on top the KamLAND energy
spectrum (not shown). For the two allowed
regions at higher values of $\delta m^2$,
such (higher-frequency) ripples are smeared away by the finite 
KamLAND energy resolution, and the contours are smooth.
More importantly, all the three
allowed regions in Fig.~8
appear to be shifted to larger values of $\sin^2\theta_{12}$, as compared with
the standard (no georeactor) case in Fig.~4. This behavior is
qualitatively  expected,
since  larger mixing is needed to suppress the excess event rate due
to the georeactor.%
%.........................................
\footnote{As a rule of thumb, a georeactor having power 
$P_\mathrm{geo}=y$ TW  increases the KamLAND rate by $\sim y\% $ 
\cite{Unbi}.}
%...........................................
For increasing $P_\mathrm{geo}$, we should then expect an increasing
tension with solar neutrino data, which fix $\sin^2\theta_{12}$ around
the value $\sim 0.3$ (as shown in Fig.~5)
independently of $P_\mathrm{geo}$.

Let us now consider the results of a maximum-likelihood analysis
where $P_\mathrm{geo}$ is free,
and the oscillation parameters are marginalized away.
The results are shown in Fig.~9, in terms of the function
$\Delta\chi^2(P_\mathrm{geo})$. From right to left, the four
curves refer to increasingly informative and powerful analyses: (1)
KamLAND likelihood in energy; (2) KamLAND likelihood in energy and time; (3)
KamLAND likelihood in energy, plus solar neutrino data; (4)
KamLAND likelihood in energy and time, plus solar neutrino data.
One can see that solar neutrino data can provide powerful
(although indirect) constraints, by forbidding the large values
of $\sin^2\theta_{12}$ preferred for $P_\mathrm{geo}>0$. To
a lesser extent, the time information in KamLAND 
(consistent with known reactor source variations) 
also disfavor any additional constant georeactor contribution.
In all four cases, we find no statistically
significant evidence for $P_\mathrm{geo}>0$, and can thus place meaningful
upper bounds on its value. In particular,
the most complete and powerful analysis in Fig.~9 (leftmost
curve) provides the bound
$P_\mathrm{geo}\lesssim 13$~TW at $2\sigma$ (95\% C.L.), not too far from
the typical expected range of a few TW \cite{Geo1}. Basically, such bound
tells us that, at $\sim2\sigma$ level, the georeactor contribution
should not exceed twice the KamLAND  normalization
uncertainty (i.e., $\sim 13\%$).

As a final remark we add that, since 
known reactor power variations help in constraining a constant
(hypothetical) georeactor neutrino flux, they can also be expected to 
help in constraining the constant (guaranteed \cite{Fior}) geoneutrino flux
below the current analysis threshold. In other words, as 
emphasized in \cite{Mori}, a maximum likelihood analysis in both energy
{\em and\/} time should provide a powerful tool for the statistical
separation of the expected geoneutrino signal in KamLAND.
Similarly, one might try to extend the current bounds
on a (hypothetical) constant antineutrino flux from the Sun
\cite{Anti} in the energy region where reactors
provide a time-variable signal.

Summarizing, we find that
the inclusion of the (monthly-binned) time information in the
KamLAND analysis corroborates the usual interpretation of the data,
in terms of an oscillation-suppressed neutrino flux generated
from known (time-variable) reactor sources.
We find no indication
for additional constant contribution from a natural
georeactor, and place an upper limit $P_\mathrm{geo}\lesssim 13$ TW 
at 95\% C.L.\  
In any case, as emphasized in the Introduction, more refined and official 
KamLAND likelihood analyses
(including, e.g., daily data about the detector and the
reactors) will be crucial to improve and check such conclusions.

\section{Conclusions}

So far, published KamLAND data analyses have been focussed 
to the energy spectrum of neutrino events. In this work, after
checking that we can reproduce in detail the official KamLAND 
likelihood analysis in energy, we have 
tried to add the time information to the analysis.
In particular, by including
monthly-binned data on Japanese reactor powers,
KamLAND event-by-event energies,
and detector livetimes, we find that the case of
actual time variations of reactor powers is significantly preferred
($\sim 3\sigma$) over the hypothetical case of no time variations. 
This interesting indication is basically unaltered by adding solar neutrino
constraints on the oscillation parameters. We have also considered
the effect of a hypothetical georeactor with power $P_\mathrm{geo}$ in
the analysis. We find increasingly tighter upper bounds as more data 
(from time variations and from solar neutrinos) are included, down
to $P_\mathrm{geo}\lesssim 13$ TW at 95\% C.L.\  

Our analysis supports
the standard interpretation of the observed KamLAND
neutrino events as generated by
known reactors sources  
and affected by flavor
oscillations with mass-mixing parameters consistent with solar
neutrino data. Implications for a hypothetical
constant neutrino source in the Earth 
interior (the georeactor) start to emerge. Other constant-flux
sources (e.g., geoneutrinos or solar antineutrinos) might
be usefully constrained in a similar way.
We hope that these encouraging results may motivate
other independent analyses of the reactor information in the
time domain, especially by the KamLAND collaboration that, by using
the complete and fully controlled data set, can certainly 
provide more reliable results and explore
interesting new facets of the topics touched in this work.

\acknowledgments

This work is supported by the
Italian Ministero dell'Istruzione, Universit\`a e Ricerca (MIUR) and
Istituto Nazionale di Fisica Nucleare (INFN) through the
``Astroparticle Physics'' research project. We thank
K.\ Inoue and the KamLAND Collaboration for permission to use
monthly-binned events and detector livetimes. We also thank
A.\ Marrone and D.\ Montanino for useful comments on the manuscript.
E.L.\ acknowledges interesting discussions on
the georeactor hypothesis with J.G.\ Learned.

\newpage
%---------------------------------------------------------------------------
\begin{figure}
\vspace*{-0cm}\hspace*{-0cm}
\includegraphics[scale=1.0]{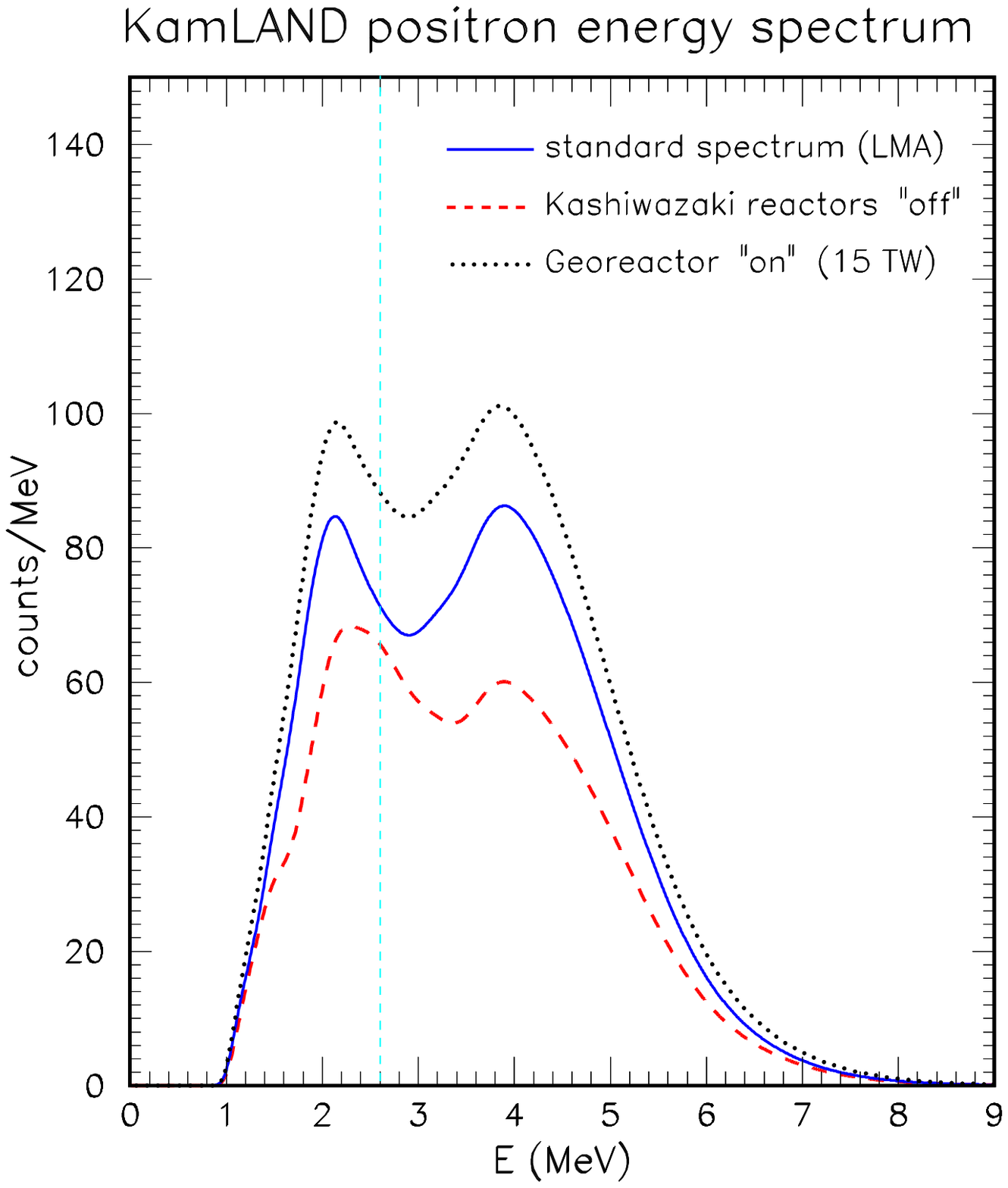} 
\vspace*{0.5cm} \caption{\label{fig1} KamLAND
absolute spectrum (without backgrounds) 
as a function of prompt energy. Solid curve:
standard spectrum for best-fit LMA parameters. Dashed curve: spectrum
with no contributions from the Kashiwazaki reactor power plant. 
Dotted curve: spectrum
with additional contribution from a 15 TW georeactor.
The vertical line indicates the analysis threshold (2.6 MeV).}
\end{figure}
%---------------------------------------------------------------------------

%---------------------------------------------------------------------------
\begin{figure}
\vspace*{4cm}\hspace*{-0cm}
\includegraphics[scale=0.95]{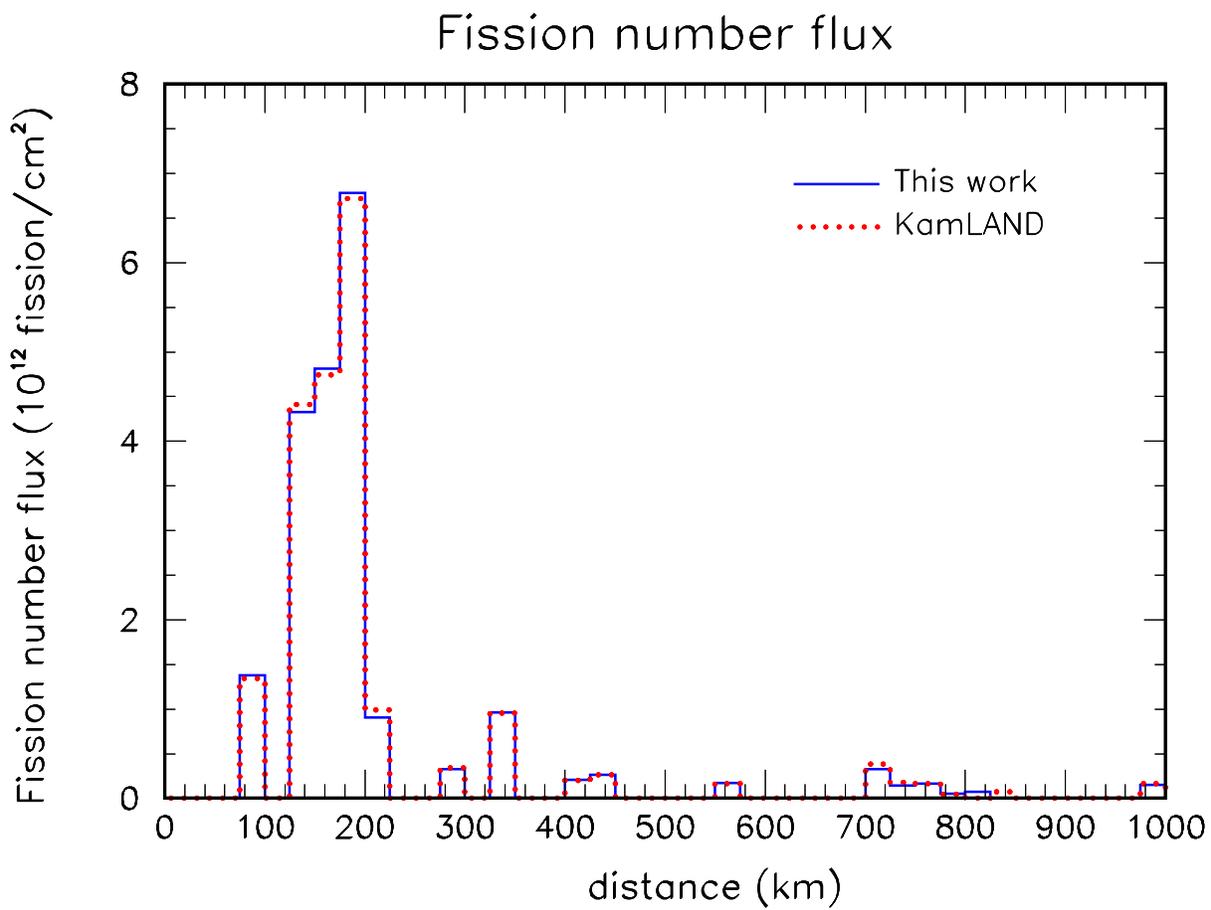} 
\vspace*{0.5cm} \caption{\label{fig2} Absolute fission number flux
at KamLAND, as a function of the reactor distance. 
Dotted histogram: KamLAND estimate. Solid histogram:
this work. }
\end{figure}
%---------------------------------------------------------------------------

%---------------------------------------------------------------------------
\begin{figure}
\vspace*{0cm}\hspace*{-0cm}
\includegraphics[scale=0.95]{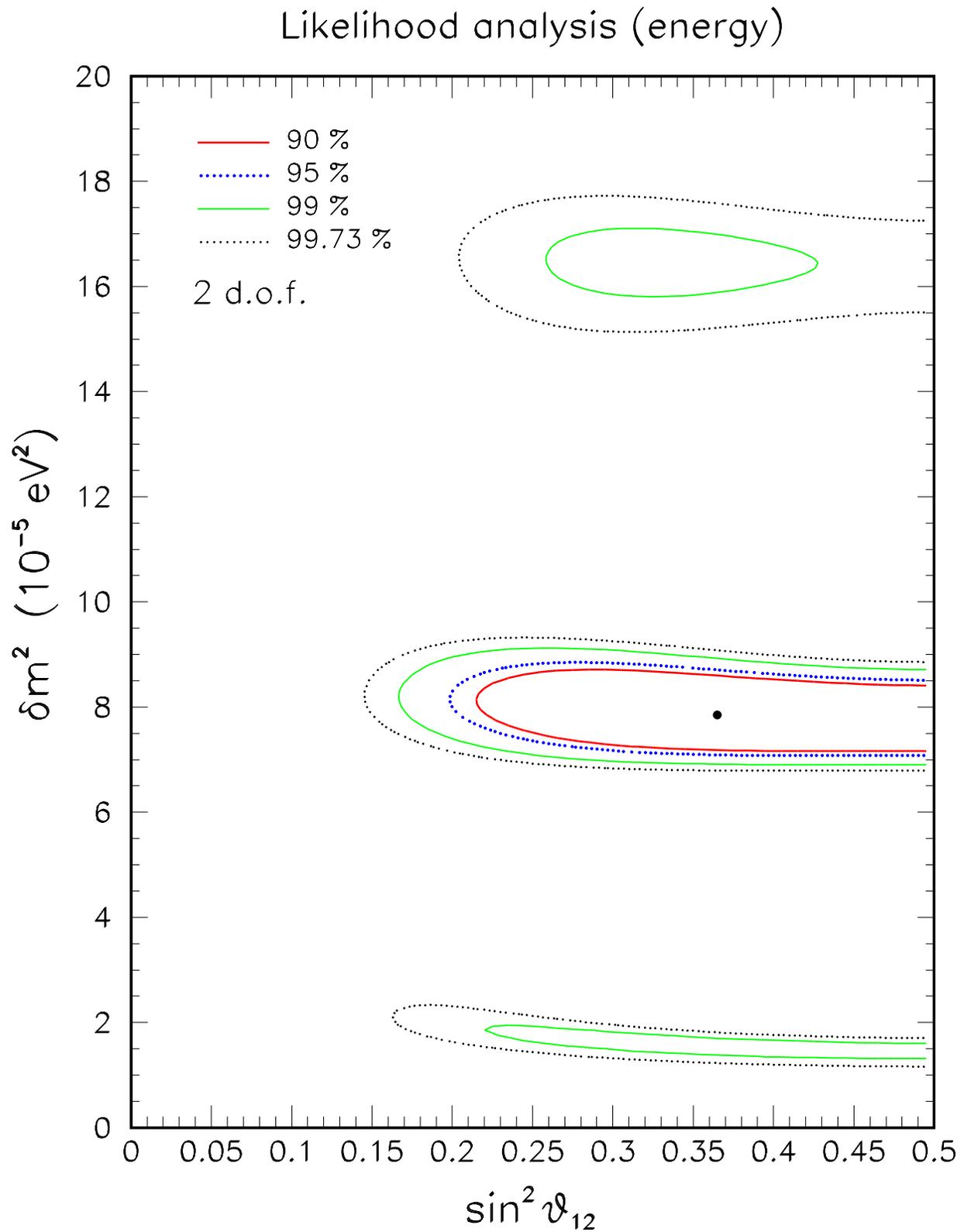} 
\vspace*{0.5cm} \caption{\label{fig3} Bounds on the oscillation parameters
from a maximum-likelihood analysis of the KamLAND energy spectrum
(event-by-event).}
\end{figure}
%---------------------------------------------------------------------------

%---------------------------------------------------------------------------
\begin{figure}
\vspace*{0cm}\hspace*{-0cm}
\includegraphics[scale=0.95]{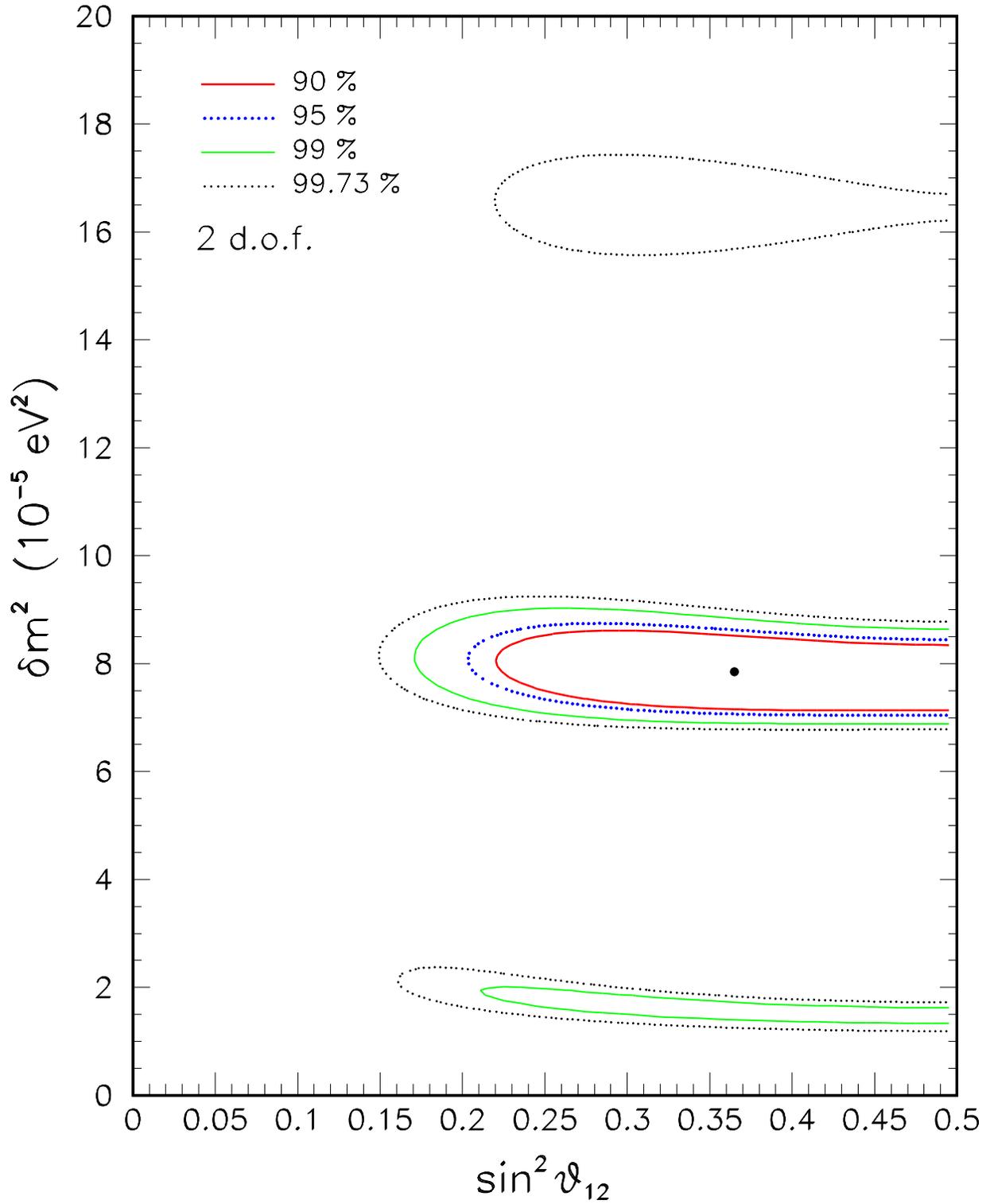} 
\vspace*{0.5cm} \caption{\label{fig4} Bounds on the oscillation parameters
from a maximum-likelihood analysis of the KamLAND information
in energy (event-by-event) and time (monthly binned).}
\end{figure}
%---------------------------------------------------------------------------

%---------------------------------------------------------------------------
\begin{figure}
\vspace*{0cm}\hspace*{-0cm}
\includegraphics[scale=0.95]{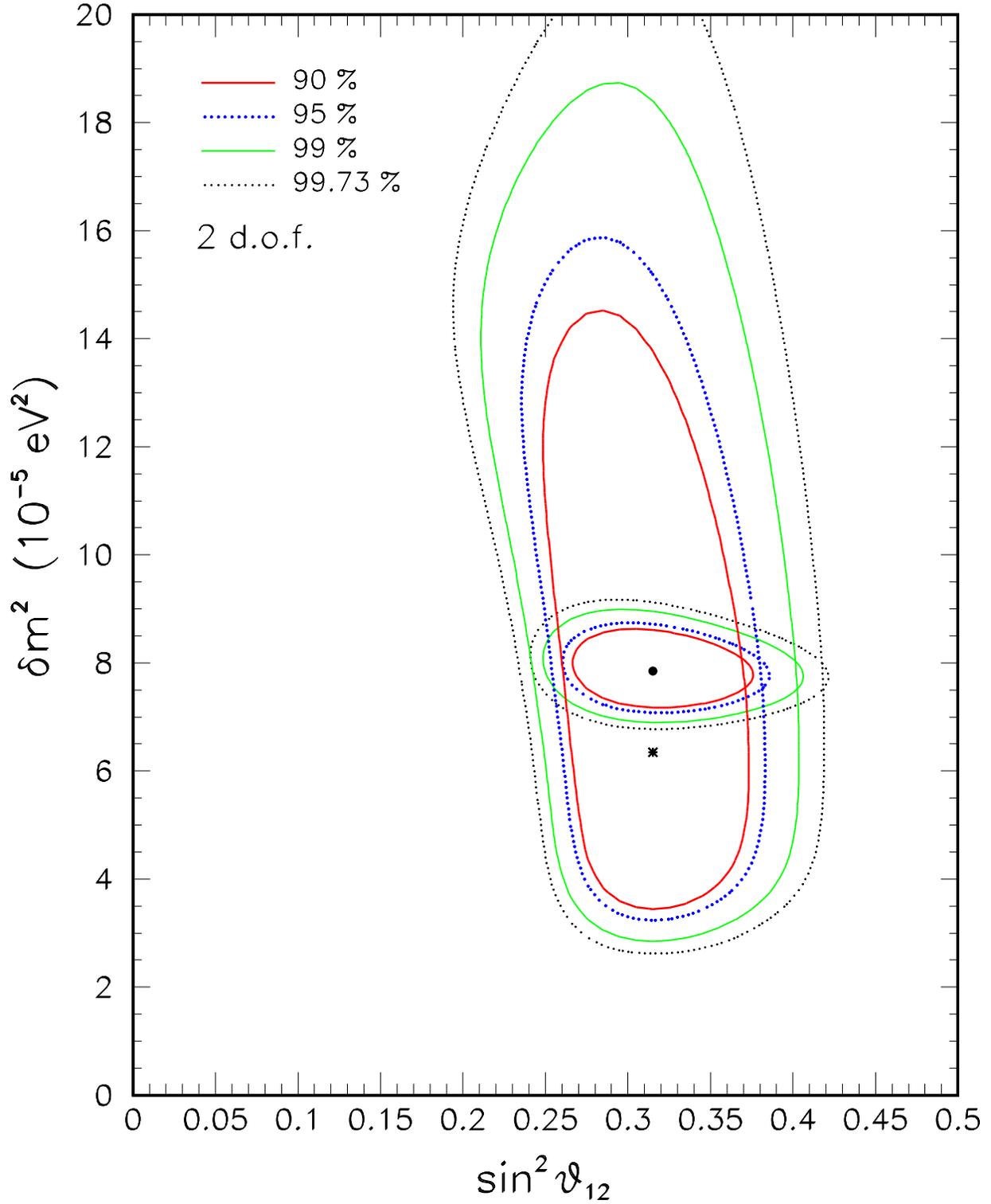} 
\vspace*{0.5cm} \caption{\label{fig5} Bounds on the oscillation parameters
from all current solar neutrino data \cite{Prog}, and their combination
with the KamLAND bounds in Fig.~4.}
\end{figure}
%---------------------------------------------------------------------------

%---------------------------------------------------------------------------
\begin{figure}
\vspace*{4cm}\hspace*{-0cm}
\includegraphics[scale=0.9]{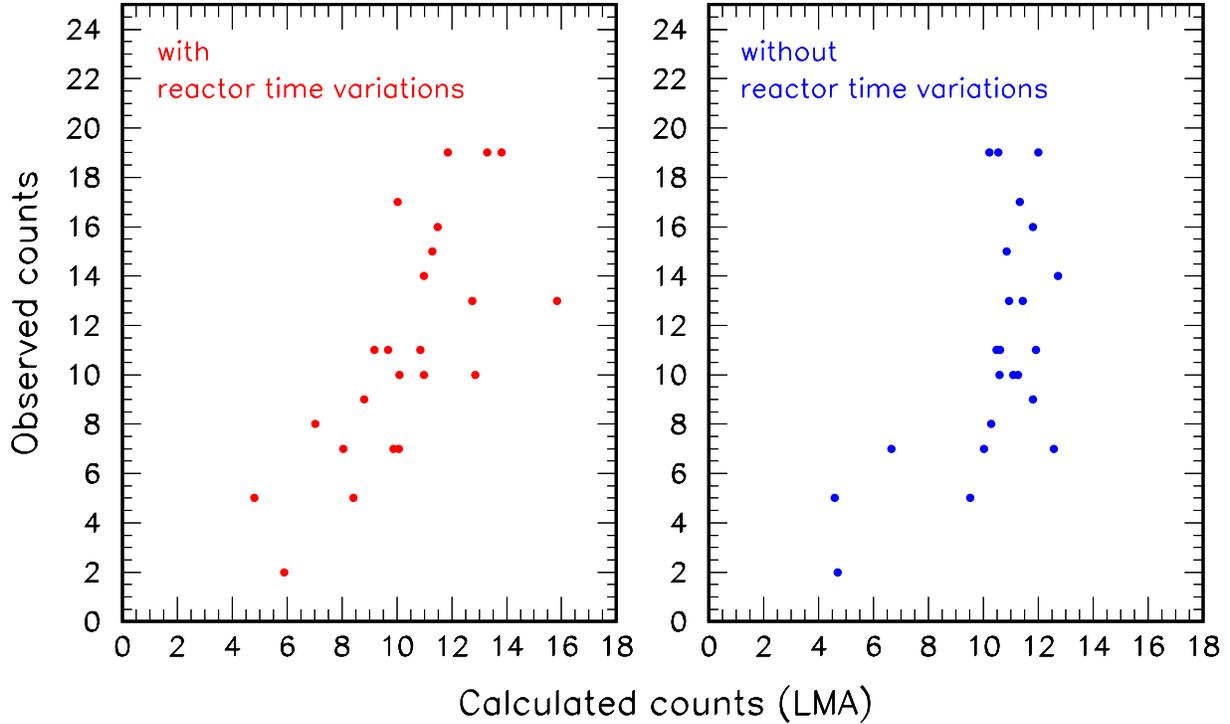} 
\vspace*{0.5cm} \caption{\label{fig6} Monthly counts of events observed in
KamLAND, plotted against the corresponding theoretical counts (calculated
for the global
best-fit LMA parameters in Fig.~5). 
Left panel: time variations of reactor
powers included.
Right panel: variations excluded (average powers used). The 
positive correlation between observed and calculated counts appears 
to be more pronounced in the first case. In both cases, monthly
KamLAND livetimes are included.}
\end{figure}
%---------------------------------------------------------------------------

%---------------------------------------------------------------------------
\begin{figure}
\vspace*{0cm}\hspace*{-0cm}
\includegraphics[scale=1.0]{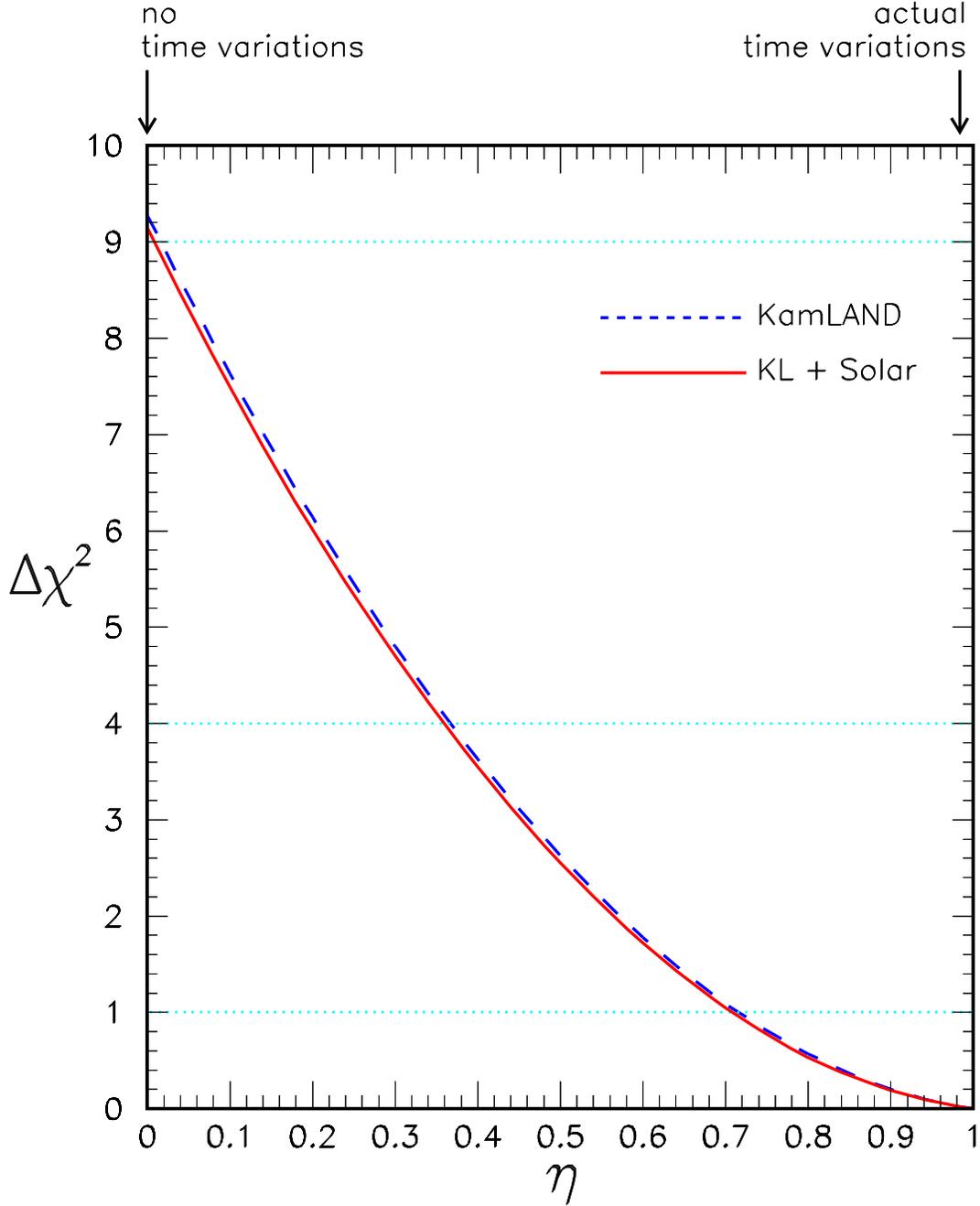} 
\vspace*{0.5cm} \caption{\label{fig7}
Bounds on the parameter $\eta$, which interpolates between the extreme
cases of no time variations of reactor powers ($\eta=0$) and actual
time variations of reactor powers ($\eta=1$). The value $\eta=1$ is 
significantly preferred over $\eta=0$.
Bounds at 1, 2, and $3\sigma$ can be obtained at $\Delta\chi^2=1$,
4, and 9 (dotted horizontal lines). The results are dominated
by the maximum likelihood analysis of KamLAND data in energy and time
(dashed line) and are not appreciably affected by including solar
neutrino data. The oscillation parameters 
$(\delta m^2,\sin^2\theta_{12})$ are marginalized in both cases.}
\end{figure}
%---------------------------------------------------------------------------

%---------------------------------------------------------------------------
\begin{figure}
\vspace*{0cm}\hspace*{-0cm}
\includegraphics[scale=0.95]{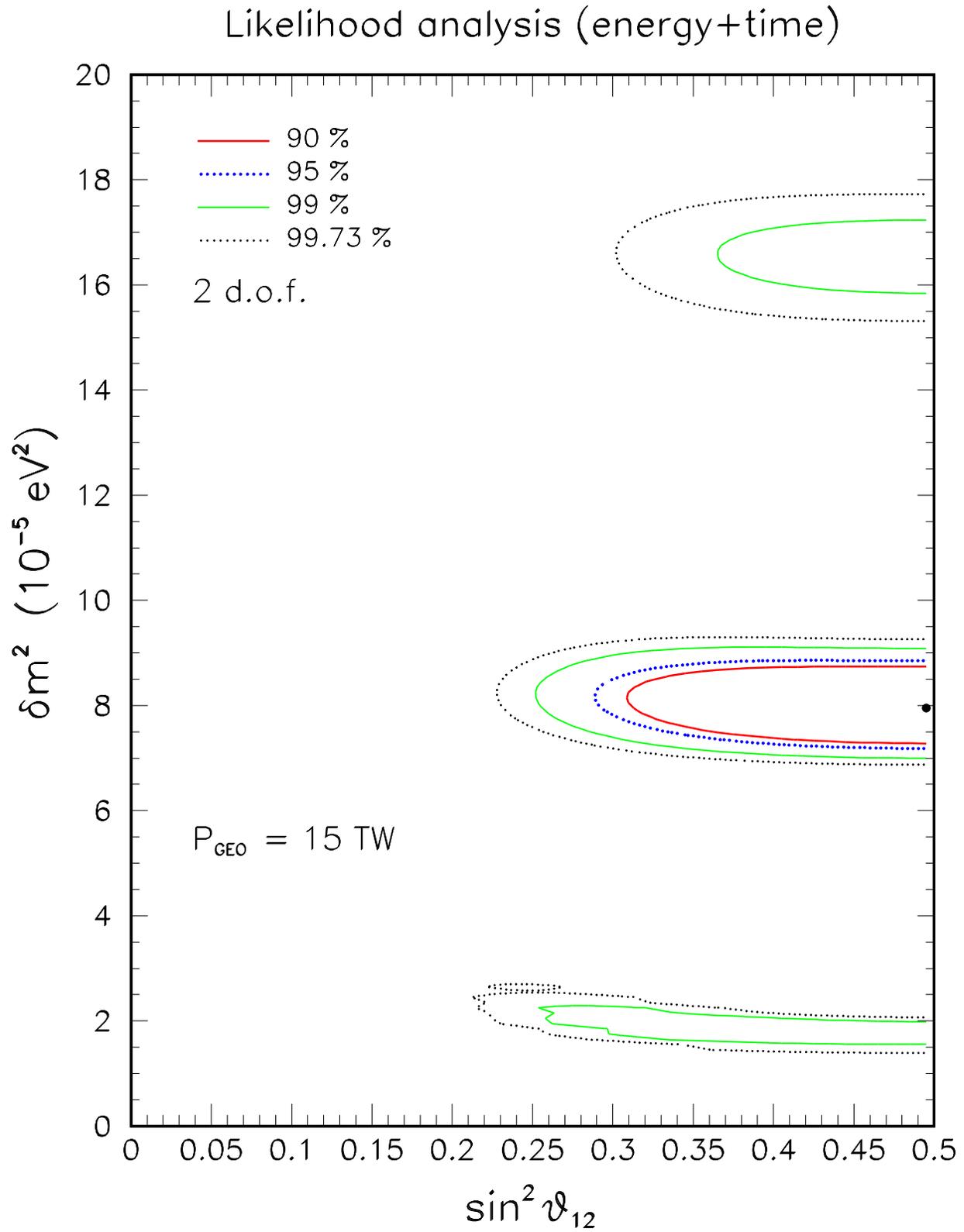} 
\vspace*{0.5cm} \caption{\label{fig8}
As in Fig.~4, but adding the contribution from a hypothetical
georeactor with $P_\mathrm{geo}=15$ TW.}
\end{figure}
%---------------------------------------------------------------------------

%---------------------------------------------------------------------------
\begin{figure}
\vspace*{0cm}\hspace*{-0cm}
\includegraphics[scale=0.95]{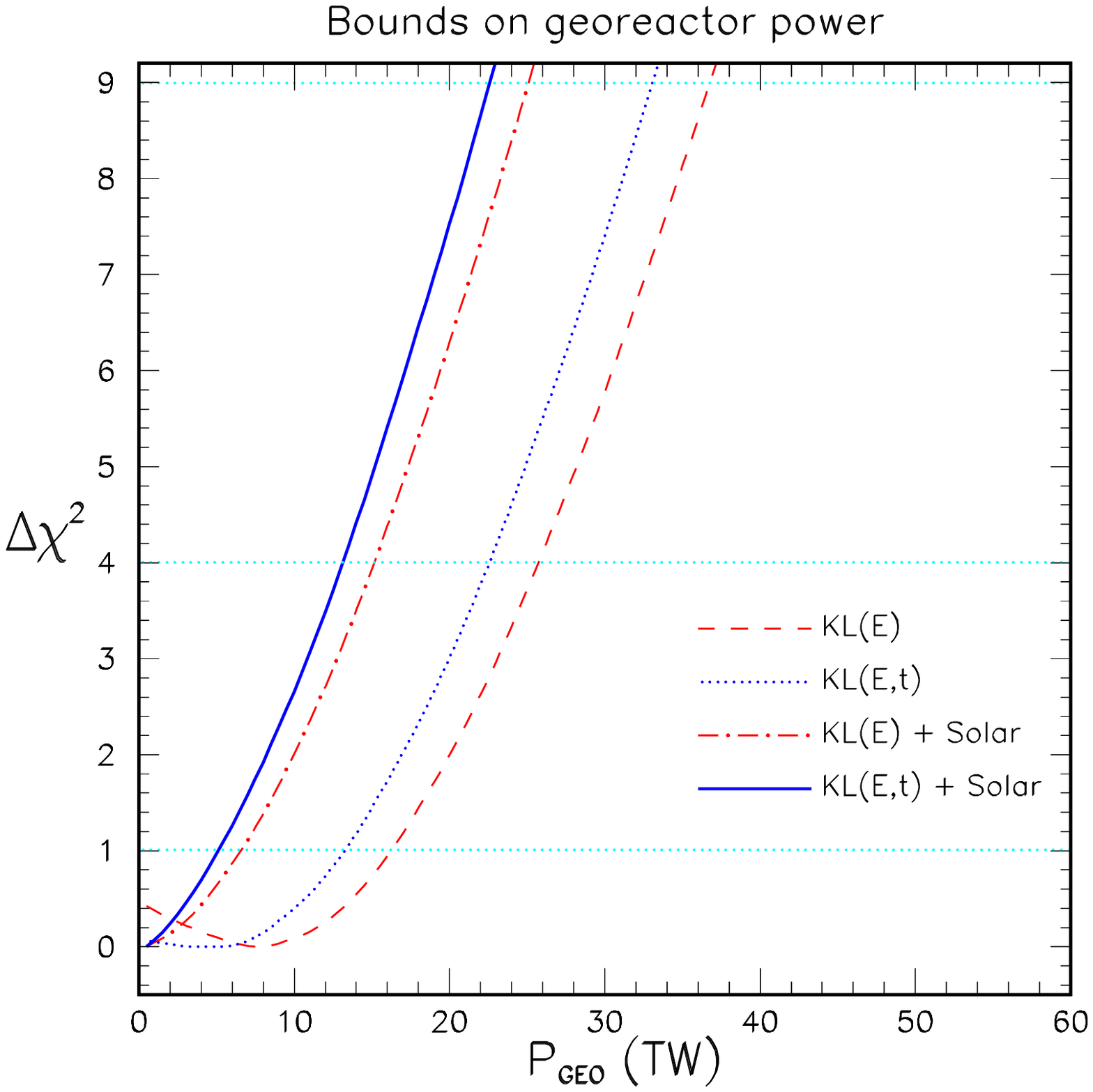} 
\vspace*{0.5cm} \caption{\label{fig9}
Bounds on the georeactor power from increasingly detailed analyses
(with marginalized oscillation parameters).
From right to left, the $\Delta\chi^2$ curves refer to: KamLAND
analysis in energy; KamLAND analysis in energy and time; 
KamLAND analysis in energy plus solar neutrino data;
KamLAND analysis in energy and time, plus solar neutrino data.
At 95\% C.L.\ ($2\sigma$), the strongest upper
bound (leftmost curve) is $P_\mathrm{geo}\lesssim 13$ TW.}
\end{figure}
%---------------------------------------------------------------------------


\begin{thebibliography}{99}


\bibitem{KamL}  	A.\ Suzuki, in the Proceedings of the Nobel Symposium
				on Neutrino Physics 
				(Haga Slott, Enk{\"o}ping, Sweden, 2004)
				ed.\ by L.\ Bergstr{\"o}m, O.\ Botner, 
				P.\ Carlson, P.O.\ Hulth, and T.\ Ohlsson,
				to appear in Physica Scripta (2005).
				Slides available at: www.physics.kth.se/nobel2004~.
				
\bibitem{Grat}	G.\ Gratta, talk at {\em Neutrino 2004}, 21st
				International Conference on neutrino Physics and
				Astrophysics (Paris, France, 2004). Slides 
				available at: neutrino2004.in2p3.fr~.

\bibitem{Pont}  B.~Pontecorvo, Zh.\ Eksp.\ Teor.\ Fiz.\ {\bf
                53}, 1717 (1968) [Sov.\ Phys.\ JETP {\bf 26}, 984
                (1968)]. 
                
\bibitem{Maki}  Z.~Maki, M.~Nakagawa, and S.~Sakata,
                Prog.\ Theor.\ Phys.\ {\bf 28}, 870 (1962).

\bibitem{Bemp}  C.~Bemporad, G.~Gratta, and P.~Vogel,
                Rev.\ Mod.\ Phys.\  {\bf 74}, 297 (2002).	

\bibitem{Kam1}	KamLAND Collaboration, K.~Eguchi {\it et al.},
			  	Phys.\ Rev.\ Lett.\  {\bf 90}, 021802 (2003).

\bibitem{Kam2}	KamLAND Collaboration, K.~Eguchi {\it et al.},
			  	Phys.\ Rev.\ Lett.\  {\bf 94}, 081801 (2005).	

\bibitem{Kam3}	KamLAND 2nd data release, 
				www.awa.tohoku.ac.jp/KamLAND/datarelease/2ndresult.html~.

\bibitem{Adia}  L.\ Wolfenstein,
                in {\it Neutrino~'78}, 8th International
                Conference on Neutrino Physics and Astrophysics
                (Purdue U., West Lafayette, Indiana, 1978), ed.\
                by E.C.\ Fowler (Purdue U.\ Press, 1978), p.~C3.

\bibitem{Matt}  L.~Wolfenstein,
                Phys.\ Rev.\ D {\bf 17}, 2369 (1978);
                S.P.~Mikheev and A.Yu.\ Smirnov,
                Yad.\ Fiz.\ {\bf 42}, 1441 (1985)
                [Sov.\ J.\ Nucl.\ Phys.\ {\bf 42}, 913 (1985)].

\bibitem{Bahc}	J.N.\ Bahcall,
				{\em Neutrino Astrophysics\/}
				(Cambridge University Press, Cambridge, UK, 1989).

\bibitem{Home}  Homestake Collaboration,
                B.T.~Cleveland, T.~Daily, R.~Davis Jr., J.R.~Distel,
                K.~Lande, C.K.~Lee, P.S.~Wildenhain, and
                J.~Ullman, Astrophys.\ J.\  {\bf 496}, 505 (1998).

\bibitem{SAGE}  SAGE Collaboration,
                J.N.~Abdurashitov {\it et al.},
                J.\ Exp.\ Theor.\ Phys.\  {\bf 95}, 181 (2002)
                [Zh.\ Eksp.\ Teor.\ Fiz.\  {\bf 95}, 211 (2002)].

\bibitem{GALL}  T.\ Kirsten for the GALLEX/GNO Collaboration,
                in the Proceedings of {\it Neutrino 2002}, 20th International
                Conference on Neutrino Physics and Astrophysics
                (Munich, Germany, 2002), edited by F.~von
                Feilitzsch and N.~Schmitz, Nucl.\ Phys.\ B {Proc.\
                Suppl.} {\bf 118}, 33 (2003).

\bibitem{GNOx}	GNO Collaboration, M.\ Altmann {\em et al.},
				hep-ex/0504037.

\bibitem{SKso}  SK Collaboration, 
                M.B.~Smy {\it et al.},
                Phys.\ Rev.\ D {\bf 69}, 011104 (2004).

\bibitem{SNO1}  SNO Collaboration, S.N.\ Ahmed {\em et al.},
                Phys.\ Rev.\ Lett.\  {\bf 92}, 181301 (2004).

\bibitem{SNO2}	SNO Collaboration,
				B.\ Aharmim {\em et al.},
				nucl-ex/0502021.

\bibitem{PDG4}  Review of Particle Physics,
                S. Eidelman {\em et al.}, Phys. Lett. B {\bf 592}, 1 (2004).

\bibitem{Unbi}	E.\ Lisi, A.\ Palazzo, and A.M.\ Rotunno,
				Astropart.\ Phys.\  {\bf 21}, 511 (2004).

\bibitem{Geo1}  J.M.\ Herndon, Proc.\ Natl.\ Acad.\ Sci.\ U.S.A.\
                {\bf 93}(2), 646 (1996); {\em ibidem\/}
                {\bf 100}(6), 3047 (2003); see also the website
                www.nuclearplanet.com~.

\bibitem{Ku03}	K.\ Inoue, private communication (2003).	

\bibitem{Fo03}	G.L.\ Fogli, E.\ Lisi, A.\ Marrone, D.\ Montanino,
				A.\ Palazzo, and A.M.\ Rotunno,
				Phys.\ Rev.\ D {\bf 67}, 073002 (2003).		

\bibitem{Ia03}	A.\ Ianni, 
				J.\ Phys.\ G {\bf 29}, 2107 (2003).	

\bibitem{Schw}	T.\ Schwetz,
				Phys.\ Lett.\ B	{\bf 577}, 120 (2003).

\bibitem{Ricc}	G.~Fiorentini, T.~Lasserre, M.~Lissia, 
				B.~Ricci and S.~Sch{\"o}nert,
  				Phys.\ Lett.\ B {\bf 558}, 15 (2003).

\bibitem{Smir}	P.C.\ de Holanda and A.Yu.\ Smirnov, 
				JCAP {\bf 0302}, 001 (2003).

\bibitem{Ba04}	J.N.\ Bahcall, M.C.\ Gonzalez-Garcia, and C.\ Pe{\~na}-Garay,
				JHEP {\bf 0408}, 016 (2004).

\bibitem{Va04}	M.\ Maltoni, T.\ Schwetz, M.A.\ Tortola, and J.W.F.\ Valle,
				New J.\ Physics {\bf 6}, 122 (2004).	

\bibitem{Al04}	P.\ Aliani, V.\ Antonelli, M.\ Picariello, and 
				E.\ Torrente-Lujan, Phys.\ Rev.\ D {\bf 013005} (2004).	

\bibitem{Ma05}	A.B.\ Balantekin, V.\ Barger, D.\ Marfatia, S.\ Pakvasa,
				and H.\ Yuksel, Phys.\ Lett.\ B {\bf 608}, 115 (2005).	

\bibitem{Go05}	A.\ Bandyopadhyay, S.\ Choubey, S.\ Goswami, S.T.\ Petcov,
				and D.P.\ Roy, Phys.\ Lett.\ B {\bf 608}, 115 (2005).	

\bibitem{St05}	A.\ Strumia and F.\ Vissani, hep-ph/0503246.	

\bibitem{JAIF}	Japan Atomic Industrial Forum,
				www.jaif.or.jp/english/aij/index2.html
				(Operating Records of Nuclear Power Plants
				in Japan). The information on this web page
				was free until March 2005; currently it requires
				subscription. 


\bibitem{File}	Monthly-binned KamLAND data on the detector livetime
				and on event-by-event
				energies are summarized in the file
				monthly\_sort\_energy.dat, currently available
				at \protect\cite{Kam3}.
				
\bibitem{Prop}	{\em Proposal for U.S.\ participation
				in KamLAND}, available at: kamland.lbl.gov/TalksPaper~.


\bibitem{Boeh}	F.\ Bohem and P.\ Vogel, {\em Physics of Massive
				Neutrinos\/} (Cambridge University Press, New York, 1992).	

\bibitem{Mura}	H.~Murayama and A.~Pierce,
  				Phys.\ Rev.\ D {\bf 65}, 013012 (2002).

\bibitem{Voge}	P.\ Vogel and J.\ Engel,
				Phys.\ Rev.\ D {\bf 39}, 3378 (1989).	


\bibitem{Beac}	P.\ Vogel and J.\ Beacom,
				Phys.\ Rev.\ D {\bf 60}, 053003 (1999).	
					

\bibitem{Mont}	G.L.~Fogli, E.~Lisi and D.~Montanino,
  				Phys.\ Rev.\ D {\bf 54}, 2048 (1996).

\bibitem{Berg}	B.\ Berger, in the {\em 40th Rencontres de Moriond},
				Electroweak Interactions and Unified Theories
				(La Thuile, Italy, 2005). Slides available at:
				moriond.in2p3.fr/EW/2005~.	

\bibitem{Prog}	G.L.\ Fogli, E.\ Lisi, A.\ Marrone, and A.\ Palazzo,
				work in progress.	

\bibitem{Lyon}	L.\ Lyons, ``Selecting between two hypotheses,''
				Oxford University Report No.\ OUNP-99-12.

\bibitem{Geo2}	J.M.~Herndon and D.~A.~Edgerley,  hep-ph/0501216.	

\bibitem{Fior}	See G.~Fiorentini, M.~Lissia, F.~Mantovani and R.~Vannucci,  
				hep-ph/0501111, and references therein.	

\bibitem{Ande}	D.\ Anderson, ``Energetics of the Earth and the missing 
				heat source mistery,'' available at 
				www.mantleplumes.org/Energetics.html~.

\bibitem{Poll}	H.N.\ Pollack, S.J.\ Hurter, and J.R.\ Johnson,
				Rev.\ Geophys.\ {\bf 31}, 267 (1993).

\bibitem{Hofm}	A.M.\ Hofmeister and R.E.\ Criss,
				Tectonophysics {\bf 395}, 159 (2005).

\bibitem{McDo}  W.F.~McDonough, {\em Compositional Models for the Earth's
                Core}, in ``Treatise on Geochemistry,'' Vol.~II,
                edited by R.W.~Carlson (Elsevier-Pergamon, Oxford, 2003); 		
                website:
                www.treatiseongeochemistry.com~.
	
\bibitem{Geos}	See, e.g., R.S.\ Raghavan, hep-ex/0208038; 
				R.J.\ de Meijer, E.R.\ van der Graaf, and K.P.\ Jungmann,
				physics/0404046; G.\ Domogatski,
				V.\ Kopeikin, L.\ Mikaelian, and V.\ Sinev, hep-ph/0407148.
				
\bibitem{APSR}	American Physical
				Society, Multidivisional Neutrino Study:
				Report of the Reactor Working Group (2004), available at 
				www.aps.org/neutrino~.

\bibitem{Mori}	G.L.~Fogli, E.~Lisi, A.~Palazzo and A.~M.~Rotunno,  
				hep-ph/0405139.

\bibitem{Anti}	KamLAND Collaboration,
				K.\ Eguchi {\em et al.}, 
				Phys.\ Rev.\ Lett.\ {\bf 92}, 071301 (2004).


\end{thebibliography}
\end{document}